\documentclass[preprint]{revtex4-1}
\usepackage[T1]{fontenc}
\usepackage{graphicx}
\usepackage{rotating}
\usepackage{bm}        
\usepackage{amssymb}   
\usepackage{bm}
\usepackage{float}
\usepackage{tikz}
\usepackage{diagbox}
\usepackage{braket}
\usepackage{color} 
\usepackage{colortbl}
\usepackage{amsmath}

\makeatletter

    \def\CT@@do@color{%
      \global\let\CT@do@color\relax
            \@tempdima\wd\z@
            \advance\@tempdima\@tempdimb
            \advance\@tempdima\@tempdimc
    \advance\@tempdimb\tabcolsep
    \advance\@tempdimc\tabcolsep
    \advance\@tempdima2\tabcolsep
            \kern-\@tempdimb
            \leaders\vrule
                    \hskip\@tempdima\@plus  1fill
            \kern-\@tempdimc
            \hskip-\wd\z@ \@plus -1fill }
    \makeatother

\def\k1{k_1}
\def\k2{k_2}
\def\q1{q_1}
\def\q2{q_2}

\def\({\left (}
\def\){\right )}
\def\[{\left [}
\def\]{\right ]}

\newcommand{\beq}{\begin{equation}}
\newcommand{\eeq}{\end{equation}}

\DeclareMathAlphabet\mathbfcal{OMS}{cmsy}{b}{n}

\begin{document}

\date{\today}
\flushbottom \draft
\title{
Neural network Gaussian processes as efficient models of potential energy surfaces for polyatomic molecules
}
\author{J. Dai}
\author{R. V. Krems}
\affiliation{
Department of Chemistry, University of British Columbia, Vancouver, B.C. V6T 1Z1, Canada \\
Stewart Blusson Quantum Matter Institute, Vancouver, B.C. V6T 1Z4, Canada}

\begin{abstract}

Kernel models of potential energy surfaces (PES)  for polyatomic molecules are often restricted by a specific choice of the kernel function.
This can be avoided by optimizing the complexity of the kernel function. For regression problems with very expensive data, the functional form of the model kernels 
can be optimized in the Gaussian process (GP) setting through compositional function search guided by the Bayesian information criterion. However, the compositional kernel search is computationally demanding and relies on greedy strategies, which may yield sub-optimal kernels. 
An alternative strategy of increasing complexity of GP kernels treats a GP as a Bayesian neural network (NN) with a variable number of hidden layers, which yields NNGP models.  Here, we present a direct comparison of GP models with composite kernels and NNGP models for applications aiming at the construction of global 
PES for polyatomic molecules. We show that NNGP models of PES can be trained much more efficiently and yield better generalization accuracy without relying on any specific form of the kernel function. 
We illustrate that NNGP models trained by distributions of energy points at low energies produce accurate predictions of PES at high energies. We also illustrate that NNGP models can extrapolate in the input variable space by building the free energy surface of the Heisenberg model trained in the paramagnetic phase and validated in the ferromagnetic phase. By construction, composite kernels yield more accurate models than kernels with a fixed functional form. Therefore, by illustrating that NNGP models outperform GP models with composite kernels, 
our work suggests that NNGP models should be a preferred choice of kernel models for PES.

\end{abstract}

\maketitle

\section{Introduction}

Simulations of quantum dynamics of polyatomic molecules require accurate models of global potential energy surfaces (PES) usually obtained by fitting the results of electronic structure calculations.   A major thrust of recent research has been to explore applications of machine learning (ML)
methods for building accurate and data-efficient models of PES. ML models of PES demonstrated in the literature are either artificial neural networks (NN) \cite{NNs-for-PES,NNs-for-PESa,NNs-for-PESb,NNs-for-PES-1a,NNs-for-PES-1b,NNs-for-PES-1c,NNs-for-PES-2,NNs-for-PES-3,NNs-for-PES-4,NNs-for-PES-5,NNs-for-PES-6,ML-for-PES,ML-for-PES2,ML-for-PES3} or kernel models \cite{general-fitting-2, BML,rabitz-1,rabitz-2,rabitz-3,kernel-for-PES,mypaper,gp-for-PES-11,gp-for-PES-2,gp-for-PES-3,gp-for-PES-4,gp-for-PES-5,gp-for-PES-6,gp-for-PES-7,gp-for-PES-8,gp-for-PES-9,gp-for-PES-10,gp-for-PES-12,gp-for-PES-13,gp-for-PES-14,gp-for-PES-15,kasrapaper,jie-jpb,hiroki,sparse-GP-PES,jie-cui-prl,gdml, gdml-1,gdml-2,sgdml}, most often in the form of kernel ridge regression (KRR) \cite{general-fitting-2, BML,rabitz-1,rabitz-2,rabitz-3,kernel-for-PES} or Gaussian process  (GP) regression \cite{mypaper,gp-for-PES-11,gp-for-PES-2,gp-for-PES-3,gp-for-PES-4,gp-for-PES-5,gp-for-PES-6,gp-for-PES-7,gp-for-PES-8,gp-for-PES-9,gp-for-PES-10,gp-for-PES-12,gp-for-PES-13,gp-for-PES-14,gp-for-PES-15,kasrapaper,jie-jpb,hiroki,sparse-GP-PES,jie-cui-prl}. Generally, the accuracy of a ML  model of PES can be improved by: (i) increasing the number $n$ of potential energy points used for training the model; (ii) optimizing the complexity of the model. For applications, where potential energy calculations are extremely time-consuming, the goal is to optimize the model in order to achieve high accuracy of PES with a given, small $n$.

For example, it has been demonstrated that the efficiency and accuracy of GP models can be systematically improved by increasing the complexity of GP kernels through compositional search that maximizes Bayesian information criterion (BIC) \cite{bic}, which approximates marginal likelihood. GP models with composite kernels can produce accurate PES at high energies using information about potential energy at low energies \cite{mypaper,hiroki} and yield global PES with up to 51 dimensions \cite{hiroki}, or up to 57 dimensions when trained simultaneously by energies and energy gradients \cite{gdml, gdml-1,gdml-2,sgdml,kasrapaper}.  However, optimizing kernels of kernel models generally requires iterative inversion of the kernel matrix, which scales as ${\cal O}(n^3)$ with the number $n$ of training points and makes compositional kernel search computationally expensive. Although this scaling of the computation complexity can be reduced by various techniques, such as data sparsification \cite{sparse-GP-PES}, optimization of sampling of energy points in the relevant configuration space \cite{mypaper,hiroki,gp-for-PES-15,gp-for-PES-14,gp-for-PES-13,gp-for-PES-12,gp-for-PES-11,gp-for-PES-2,gp-for-PES-5,gp-for-PES-8} or building molecular symmetries into ML model kernels \cite{sgdml,gdml-2}, it remains a major obstacle for applications of kernel models to high-dimensional PES.

NNs do not suffer from the ${\cal O}(n^3)$ scaling problem and can often be trained much more efficiently than accurate kernel models. However, 
can NNs provide more accurate models of PES than typical kernel models given the same number of potential energies? There is no clear answer to this question. The one study that directly compared GP regression with NN models of PES used for the calculations of ro-vibrational energy levels \cite{gp-for-PES-4} indicated that GP models yield more accurate results with fewer energy points. However, neither the hyperparameters of NNs nor the functional form of the kernels of GP regression were fully optimized in Ref. \cite{gp-for-PES-4}. Previous work indicates that KRR and GP models are competitive with the NN models in terms of accuracy and data efficiency of PES models for polyatomic molecules \cite{general-fitting-2, BML,rabitz-1,rabitz-2,rabitz-3,kernel-for-PES,mypaper,gp-for-PES-11,gp-for-PES-2,gp-for-PES-3,gp-for-PES-4,gp-for-PES-5,gp-for-PES-6,gp-for-PES-7,gp-for-PES-8,gp-for-PES-9,gp-for-PES-10,gp-for-PES-12,gp-for-PES-13,gp-for-PES-14,gp-for-PES-15,kasrapaper,jie-jpb,hiroki,sparse-GP-PES,jie-cui-prl,gdml, gdml-1,gdml-2}. 
At the same time, unlike conventional NN and KRR, GP regression offers not only models of PES, but also the Bayesian model uncertainty, which can be exploited for applications such as Bayesian optimization \cite{BML,bo,bo2,gp-for-PES-14} or the analysis of the effects of PES uncertainties on the accuracy of calculations of molecular observables \cite{jie-cui-prl}.

Since the original work on GP regression \cite{neal1}, it has been known that GP models can be viewed as (Bayesian) NNs  with infinite width. 
This connection has recently been elucidated \cite{nngp,nngp1,nngp2} and exploited \cite{nngp3,nngp4} to develop NNGP models \cite{nngp,nngp1,nngp2,nngp3,nngp4}. NNGP models 
can combine multiple layers of Bayesian NN with different properties, yielding an algorithm to increase the complexity of a GP kernel. 
At the same time, NNGP models can potentially be trained much more efficiently than kernel models with complex, composite kernels.  
However, the application of NNGP models for fitting PES has not been explored. 
In particular, it is not known if NNGP models can extrapolate as well as interpolate target functions based on small data sets. The ability of a model to extrapolate is important for applications aiming to build high-dimensional PES with a small number of energy points, where the training point distributions are likely to miss some important parts of the PES landscape.

In this work, we present a direct comparison of NNGP models and GP models based on composite kernels with optimized functional form. We build the global  PES for a polyatomic molecule with six degrees of freedom by both interpolation and extrapolation in the energy domain (i.e. build PES at high energies using energy points at low energies). We illustrate that, in both cases, NNGP models yield more accurate results than GPs with composite kernels. At the same time, we show that training NNGP models is much more efficient than training GP models with composite kernels.  
We analyze the eigenvalue decomposition of the GP model kernels and show that the kernels of NNGP models align better with the target PES than the composite kernels, indicating better learning ability of NNGP  kernels. 
To further illustrate the generalization power of NNGP regression, we use NNGP models to build the free energy surface for the Heisenberg spin model. The results illustrate that NNGP models can predict the onset of a phase transition by extrapolation of the free energy surface within one given phase. 
The results of this work illustrate that GP regression and NN modelling can be combined to yield efficient, yet accurate PES with a small number of energy points $n$.

\section{Gaussian process models with complex kernels}

A PES is represented by a function $y = f(\bm x)$, where  $\bm{x} = [x_1,x_2, ...,x_p]^{\top}$ is a vector of variables describing a molecule with $p$ degrees of freedom. A supervised ML model of PES aims to infer this function from a finite set of potential energy points $\bm{y} = [y_1,y_2,...,y_n]^{\top}$ corresponding to 
$n$ vectors $\bm x$ collected in the rectangular $n \times p$ matrix $\bm{X}$. 

 Multiple previous articles have modelled PES by Gaussian processes (GPs) \cite{mypaper,gp-for-PES-11,gp-for-PES-2,gp-for-PES-3,gp-for-PES-4,gp-for-PES-5,gp-for-PES-6,gp-for-PES-7,gp-for-PES-8,gp-for-PES-9,gp-for-PES-10,gp-for-PES-12,gp-for-PES-13,gp-for-PES-14,gp-for-PES-15,kasrapaper,jie-jpb,hiroki,sparse-GP-PES,jie-cui-prl}. A GP can be viewed as an infinite-dimensional multivariate normal distribution $\bm{Y} = [Y_1,Y_2,...,Y_\infty]^{\top} \sim \mathcal{N}({\mu}, {\bf \Sigma})$, with $Y_i$ representing the output of a GP at a particular point of input space $\bm x_i$. The distribution is infinite-dimensional because each component of $\bm x$ is a continuous variable. Therefore, a GP is entirely determined by a mean function $ \mu(\bm x)$ and a covariance function $\Sigma(\bm x, \bm x')$. The covariance function of a GP is represented by the kernel function $k(\bm x, \bm x')$. Training a GP amounts to conditioning $\bm Y$ by the given energy points $\bm y$, which changes the mean  and covariance functions of the GP. In the present work, we distinguish two types of GP models: GP models with composite kernels and Neural Network Gaussian Process (NNGP) models.

\begin{figure}[http]
\centering
\includegraphics[scale=1]{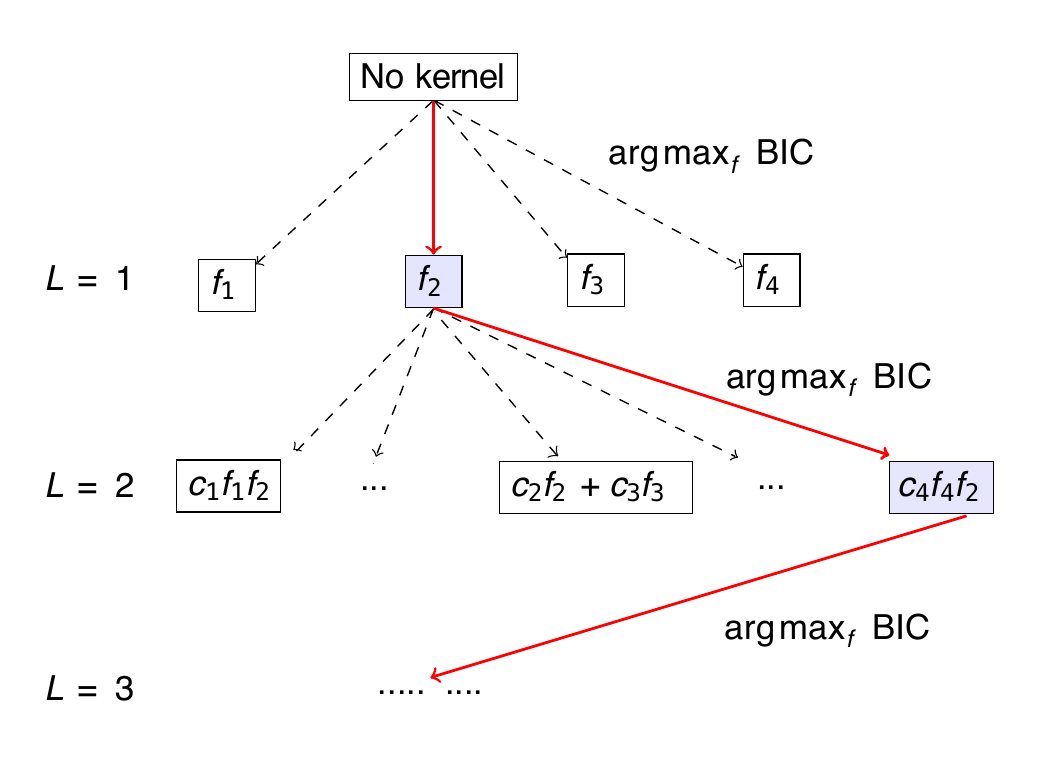} 
\caption{Schematic diagram of the algorithm for the construction of composite kernels for GP models.  Each function $f_i$ with $i \in [1, \dots,4]$ represents a parametrized kernel function. The algorithm uses Bayesian information criterion to select at each layer $L$ either a product or a linear combination of $f_i$ with the kernel function selected at the previous layer. }
\label{fig:kernel-tree}
\end{figure}

As described below, both of these algorithms allow for kernels of varying function complexity. In both cases, the complexity of the kernel function can be specified by the number of layers in the kernel construction algorithm. We denote the number of layers in the kernel construction algorithm by $L$, for both types of models. It is important to note, however, that $L$ has a different meaning for the two types of models. As explained below, $L$ denotes the depth of the search tree for the composite kernels and the number of NN layers for the NNGP models.

As was shown previously \cite{mypaper, hiroki, kasrapaper}, the mathematical form of the covariance function can be optimized to yield accurate PES with a small number of energy points.  Introduced by Duvenaud {et al} \cite{extrapolation-1,extrapolation-2} and applied to building accurate PES in Refs. \cite{mypaper, hiroki, kasrapaper}, this algorithm amounts to search in an infinite space of kernel functions. To make the search feasible, Duvenaud  et al used a greedy strategy to grow incrementally the complexity of the kernel function \cite{extrapolation-1,extrapolation-2}. The algorithm combines a set of simple kernel functions $f_i$ with $i \in [1, \dots, N_k]$ into products and linear combinations by an iterative process, choosing the most optimal kernel function at each layer of kernel complexity.  Here, $N_k$ is the number of kernel functions $f_i$ considered.
Each function $f_i$ is chosen to have properties of kernel functions of a reproducing kernel Hilbert space, which guarantees that the resulting complex function can be used as the kernel function for a kernel ML model. 
As was previously demonstrated, this kernel construction algorithm yields GP models that produce accurate PES by both interpolation and extrapolation \cite{mypaper,hiroki,kasrapaper,qgp}. 
However, the iterative optimization of kernel complexity for identifying optimal kernel functions for a given PES is numerically expensive \cite{mypaper, hiroki, kasrapaper,sparse-GP-PES}. 
As illustrated in Figure \ref{fig:kernel-tree}, the iterative algorithm of Duvenaud  et al can be mapped on a search tree, with the complexity of the kernel function increasing with the depth level ($L$) of the tree. 
The computational complexity arises from the requirement to train multiple GP models for a given depth layer $L$ and the increasing number of kernel parameters as $L$ increases. 
In the present work, we show that accurate GP models of PES can be obtained using an alternative method for increasing kernel complexity, exploiting the connection between GPs and NNs. We show that the resulting NNGP models can be trained much faster than GP models with composite kernels, while yielding more data-efficient PES.

A fully connected feed-forward neural network with multiple hidden layers enumerated by $l \in [1, L]$ is computed as 
\begin{equation}
y_{i}^{l}(\bm{x})=b_{i}^{l}+\sum_{j=1}^{N_{l-1}} W_{i j}^{l} z_{j}^{l-1}(\bm{x}),
\label{NN}
\end{equation}
where 
\begin{equation}
\quad z_{j}^{l-1}(\bm{x})=\phi\left( y_{j}^{l-1}(\bm{x})\right),
\end{equation}
$W_{i j}^l$ and $ b_i^l$ are the weight and bias parameters for the layer $l$, $\phi(y)$ is a non-linear activation function,  and $N_{l}$ is the number of nodes in layer $l$. 
The outputs $y^L_i$ of layer $L$ can then be collected into a linear combination to yield a scalar output $y$ for a single-output regression problem. 
For a NN with a single hidden layer $L = 1$, this reduces to 
\begin{equation}
y(\bm{x})=b^{1}+\sum_{j=1}^{N_1} W_{j}^{1} y_{j}^{1}(\bm{x}), 
\label{NN_1}
\end{equation}
with 
\begin{equation}
\quad y_{j}^{1}(\bm{x})=\phi\left( b_{j}^{0}+\sum_{i=1}^{p} W_{i j}^{0} x_i\right),
\end{equation}
where $x_i$ is the $i$th component of the $p$-dimensional vector $\bm x$.

The connection between single-layer NN and GP is provided by the Central Limit Theorem (CLT) \cite{neal1}. In the limit  of infinite width $N_1 \to \infty$ with priors $\mathcal{N}(\mu_w, {\sigma_{w}}^{2})$ and $\mathcal{N}(\mu_b, {\sigma_{b}}^{2})$ on the weight and bias parameters respectively,  the network becomes a Gaussian process, 
\begin{eqnarray}
y(\bm x) \sim \mathcal{GP}(\mu^1,K^1)
\end{eqnarray} 
with mean $\mu^1$ and covariance $K^1$ functions. The means of the priors can be chosen as $\mu_w = \mu_b = 0$, yielding $\mu^1=0$ and the covariance
\begin{equation}
 K^1(\bm{x},\bm{x'}) = \mathbb{E}\left[y\left(\bm{x}\right) y\left(\bm{x'}\right)\right] ={\sigma^{(1)}_{b}}^{2}+\sum_{j} {\sigma^{(1)}_{w}}^{2}\mathbb{E} \left[y^1_j\left(\bm{x}\right) y^1_j\left(\bm{x'}\right)\right].
\label{K_1}
\end{equation}

It was recently demonstrated that this can be extended to networks with multiple hidden layers $L > 1$ \cite{nngp,nngp1,nngp2,nngp3,ntk}. Given that $y_j^{l-1}$ is an independent Gaussian process for each $j$, as $N_l \to \infty$, $y_i^l$ is a Gaussian process $\mathcal{GP}(0,K^l)$ with the covariance

\begin{equation}
 K^l(\bm{x},\bm{x'}) = \mathbb{E}\left[y_i^l\left(\bm{x}\right) y_i^l\left(\bm{x'}\right)\right] ={\sigma^{(l)}_{b}}^{2}+ {\sigma^{(l)}_{w}}^{2}\mathbb{E}_{y_i^{l-1} \sim \mathcal{GP}(0,K^{l-1})} \left[\phi(y^{l-1}_i\left(\bm{x}\right)) \phi(y^{l-1}_i\left(\bm{x'}\right))\right].
\label{K_l}
\end{equation}
Any two $y_i^l$ and $y_j^l$ in the same layer are independent and follow a joint Gaussian distribution with zero covariance when $i \neq j$.  The recursive form of $K^l$ depends on the activation function $\phi$. In the present work, we choose the error function $\mathrm{erf}(y) = \frac{2}{\sqrt{\pi}} \int_{0}^{y} e^{-t^{2}} d t$ as the activation function,  and treat $\sigma^{(l)}_{w}$ and $\sigma^{(l)}_{b}$ as trainable parameters. The number of the parameters scales linearly with the number of layers $L$.

A kernel function satisfies the eigenvalue equation 
\begin{eqnarray}
 \int k(x,y) \phi_i(y) d y =  \eta_i \phi_i(x) ~~~~{\rm with} ~~~ \eta_i \geq 0 
\end{eqnarray}
where $\{\phi_{i}\}$ form an orthogonal set, and $\left\{\eta_{i}\right\}$ are the kernel eigenvalues.
If the target function $f(\bm x)$ is represented by the basis expansion
\begin{eqnarray}
{f}(\bm{x})=\sum_{i} \bar{w}_{i} \psi_{i}(\bm{x})
\end{eqnarray}
with $\psi_{i}(\bm{x}) \equiv \sqrt{\eta_{i}} \phi_{i}(\bm{x})$, the data efficiency of a model can be quantified by the cumulative power distribution \cite{kernel-spectral}
\begin{equation}
C(i)=\frac{\sum_{i^{\prime} \leq i} \eta_{i^{\prime}} \bar{w}_{i^{\prime}}^{2}}{\sum_{i^{\prime}} \eta_{i^{\prime}} \bar{w}_{i^{\prime}}^{2}}.
\label{C_power}
\end{equation}
The rate of convergence of the cumulative distribution (CD) to one can be used to quantify the efficiency of a kernel for a particular problem. 
Ideally, the kernel function should be chosen such that the target function is completely represented by a single eigenfunction of $k(\bm x, \bm x')$, which would yield $C(i) = 1$ with just $i =1$. 
We use CD defined in Eq. (\ref{C_power}) to compare kernel performance in this work in addition to analyzing the errors of the models.

\section{Results}

 
 We examine the generalization power of NNGP models by building the global six-dimensional (6D) PES for $\mathrm{H_3O^+}$, using the results of the {\it ab initio} calculations from Ref. \cite{h3o+}. The molecular geometry is described by a 6D input vector, with elements given by the Morse variables defined as $m_{i j}=\exp \left(-r_{i j} / b\right)$, where $r_{i j}$ is the distance between atoms $i$ and $j$, and the range parameter $b$ is fixed at 2.5. These molecular descriptors are the same as used in previous work \cite{gp-for-PES-10,mypaper}. The performance of the models of PES is characterized by the root mean squared error (RMSE)
 \begin{eqnarray}
\textrm{RMSE} = \sqrt{\frac{1}{N}\sum_{i}^{N} (\hat y_i - y_i )^2},
 \end{eqnarray}
 where $\hat y_i$ is the prediction of the model at $\bm x_i$ and the sum is over $N$ points that are not used for training the model, forming a hold-out test set of input-output pairs. 
Ref. \cite{h3o+} provides $31000$  {\it ab initio} points that span the energy range $[0, 21000]$ cm$^{-1}$.  In all calculations presented here, the hold-out test set comprises all {\it ab initio} points that are not used for training the specific model under examination.

 \subsection{Comparison of model accuracy}
 
\begin{figure}[http]
\centering
\includegraphics[scale=1]{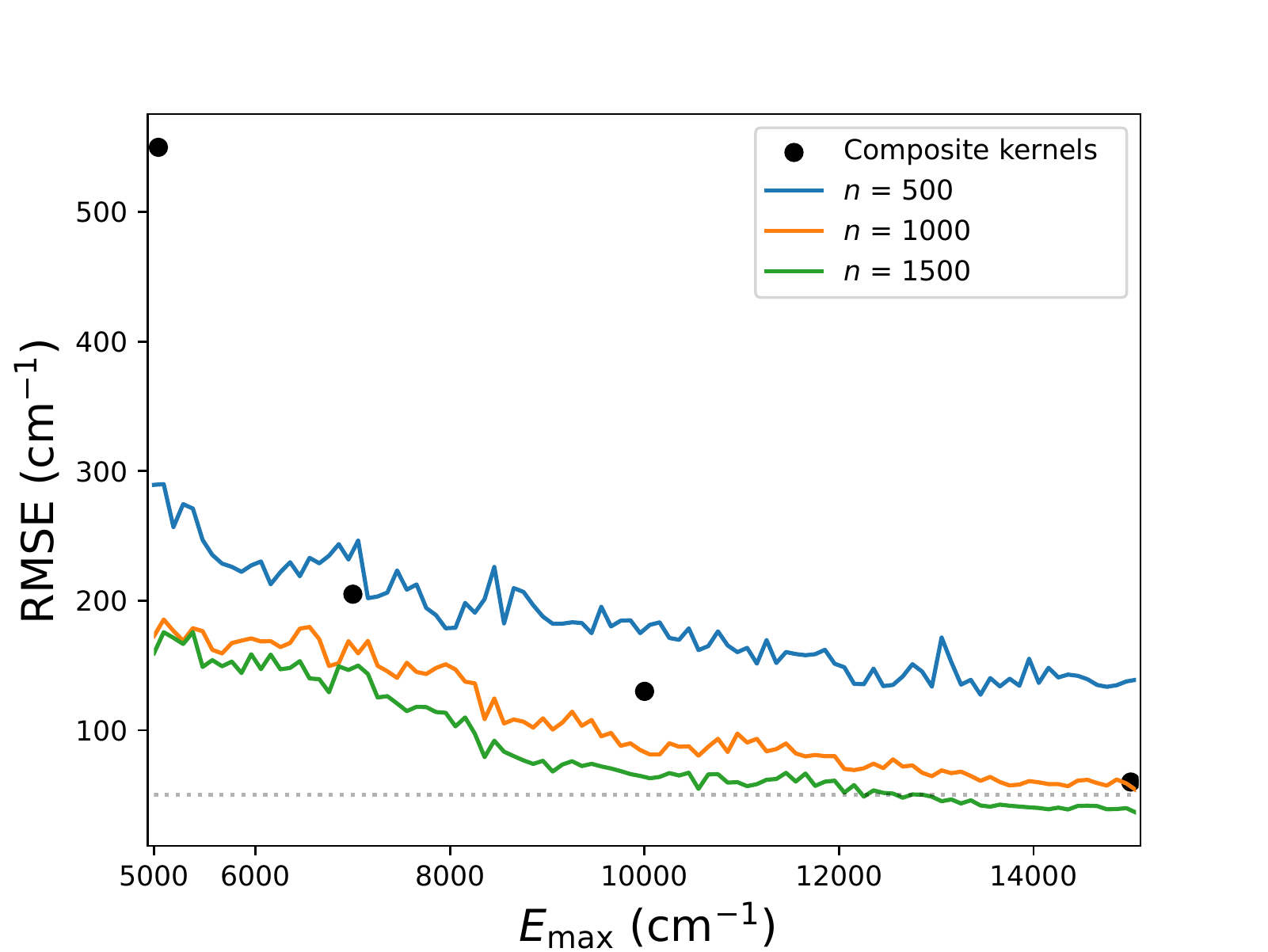} 
\caption{Global root mean squared error (RMSE) for the 6D PES for H$_3$O$^+$: solid curves --  NNGP models  trained with different numbers $n$ of potential energy points; 
symbols -- GP model with composite kernels trained by 1500 potential energy points. In each case, the training points are randomly selected from the energy range below ${E}_{\rm max}$, while the RMSE is calculated over the entire set of energy points in the  energy range up to $21,000$ cm$^{-1}$. The horizontal dotted line shows the value $50$ cm$^{-1}$. The kernels of NNGP models include $5$ layers and the composite kernels are obtained with $L = 5$ in the kernel construction algorithm. 
\\
    }
\label{fig:extrapolation}
\end{figure}

To demonstrate the extrapolation ability of the NNGP kernel, we trained a series of 6D NNGP models of PES for $\mathrm{H_3O^+}$ with 500, 1000, and 1500 training points, randomly selected from a range of energies up to a maximum of $E_{\rm max}$. 
The results are compared with the predictions of a GP model with the most accurate composite kernel trained using 1500 {\it ab initio} points randomly selected from the same range of energies.
Figure \ref{fig:extrapolation} illustrates that NNGP models trained with 1000 {\it ab initio} points outperform GP models with composite kernels trained by 1500  {\it ab initio} points. The advantage of the NNGP models is particularly noticeable for lower values of $E_{\rm max}$, which indicates that NNGP models provide more accurate extrapolation in the energy domain than the GP models with composite kernels. 

\begin{figure}[http]
\centering
\includegraphics[scale=1]{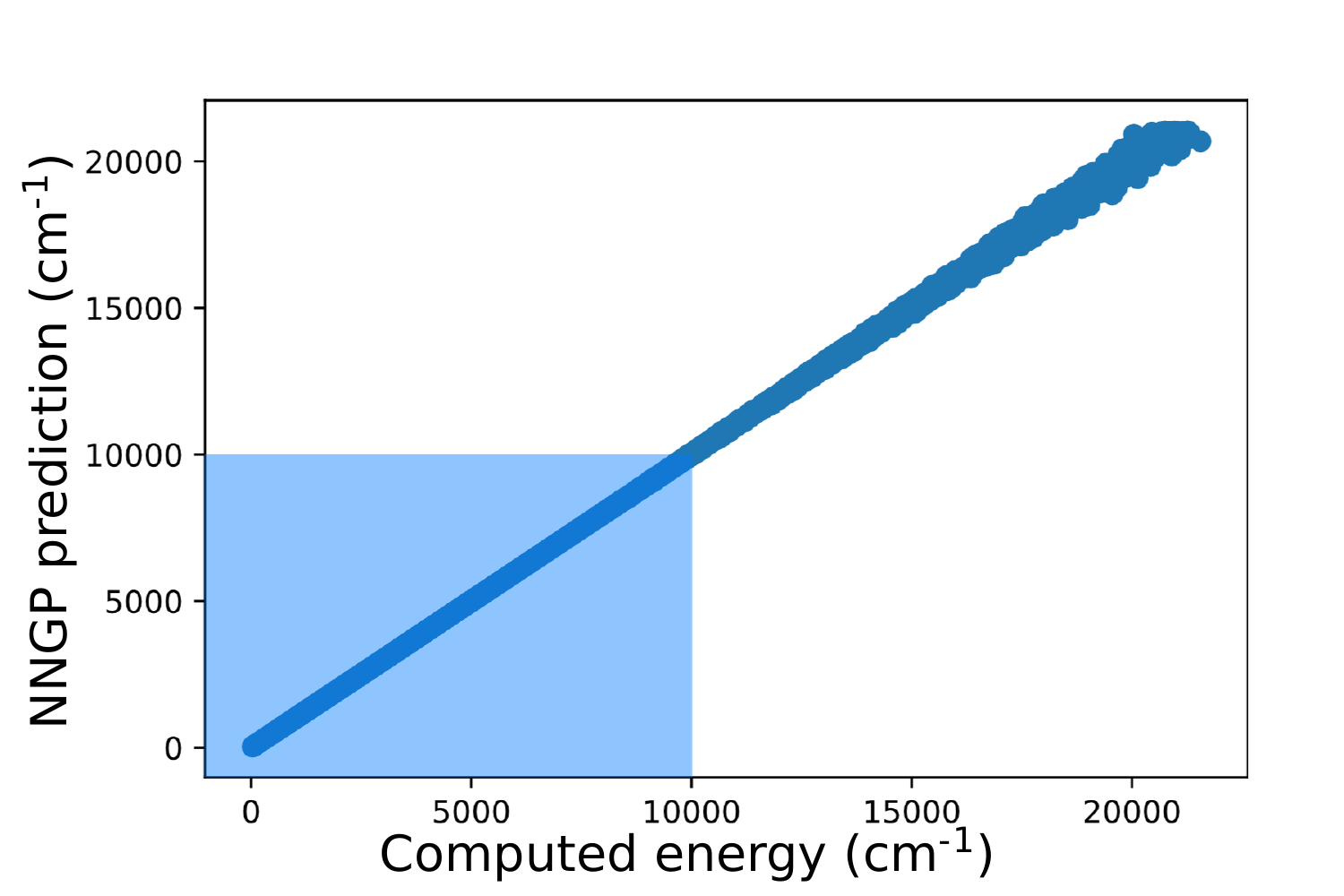} 
\caption{Comparison of the NNGP model predictions with the potential energy points computed in Ref. \cite{h3o+} for the 6D PES of H$_3$O$^{+}$.  The NNGP models are trained by a random distribution of 1500 points in the energy range $[0,10000]~\mathrm{cm^{-1}}$ shown by the shaded area.}
\label{fig:comparsion}
\end{figure}

Figure \ref{fig:comparsion} compares explicitly the predictions of the NNGP model trained by 1500 {\it ab initio} points in the energy range $[0,10000]$ cm$^{-1}$ with the computed {\it ab initio} energies in the entire energy interval $[0, 21000]$ cm$^{-1}$. 
We use NN with five layers to build the NNGP model for this calculation. The results illustrate the remarkable generalization accuracy of NNGP models, yielding accurate global 6D surface by both interpolation and extrapolation (in the energy domain) of $1500$ energy points.

\begin{figure}[http]
\centering
\includegraphics[scale=1]{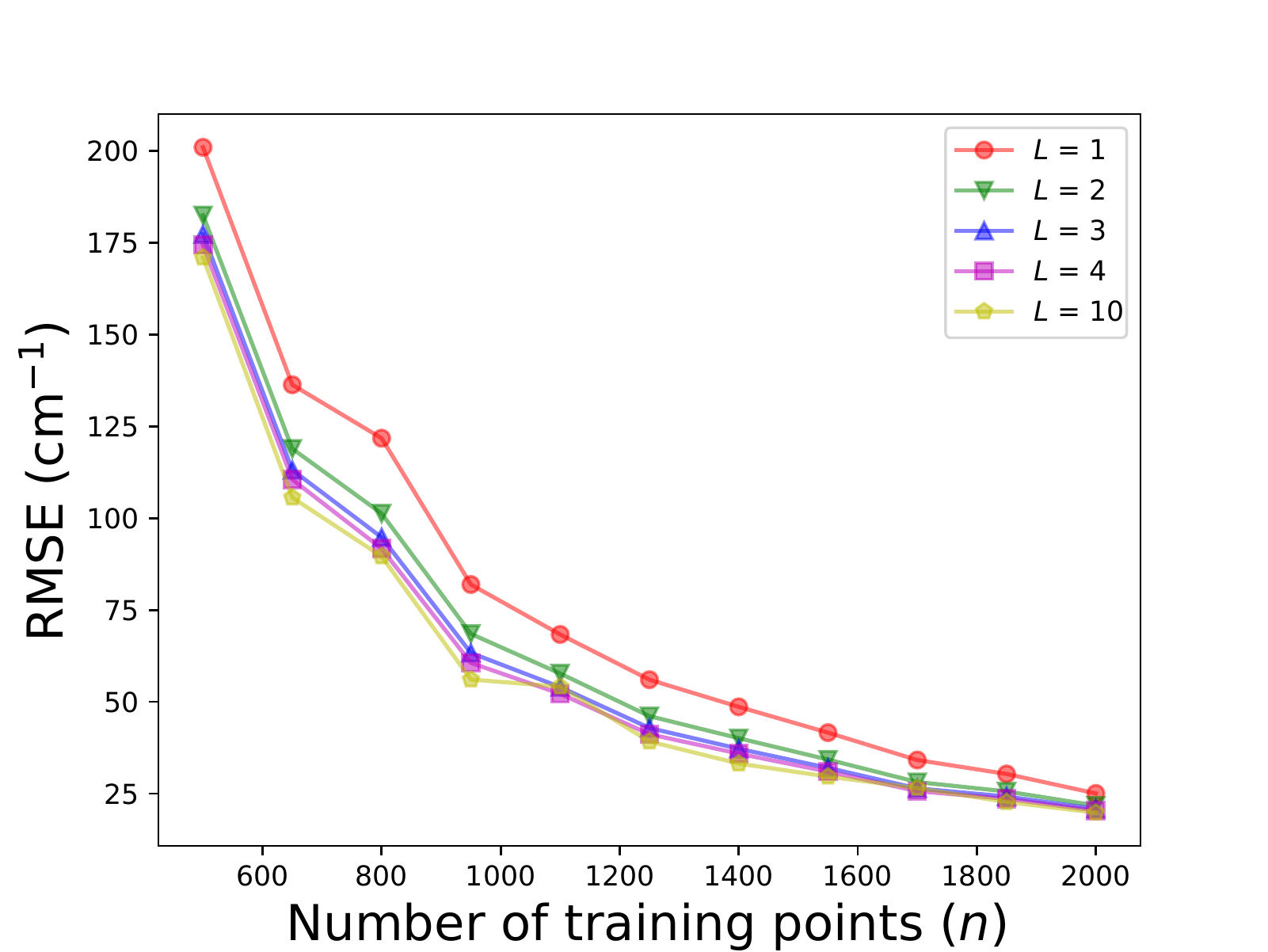} 
\caption{Root mean squared errors (RMSE) of NNGP interpolation models of the 6D PES for H$_3$O$^+$ as functions of the number of training points used to build the models. 
The training points are selected randomly from the energy interval $[0, 21000]$ cm$^{-1}$ spanning the entire PES. The different symbols correspond to the different number of NN layers used to build the kernels of NNGP models:
$L = 1$ (circles), $L = 2$ (down triangles), $L = 3$ (up triangles), $L = 4$ (squares) and $L=10$ (pentagons).}
\label{fig:interpolation}
\end{figure}

It is important to examine the convergence of NNGP models with the number of NN layers. Figure \ref{fig:interpolation} shows the RMSE for interpolation models of PES as a function of the number of training points for NNGP models with varying number of NN layers $L$. 
The calculations reveal a significant improvement of the NNGP models as the number of NN layers is increased from $L = 1$ to $L=2$. At the same time, we observe little change between predictions of NNGP models with $L =2$ and $L = 10$. 

 \subsection{Comparison of kernel eigenvalues and training complexity}

\begin{figure}[http]
\centering
\includegraphics[scale=1]{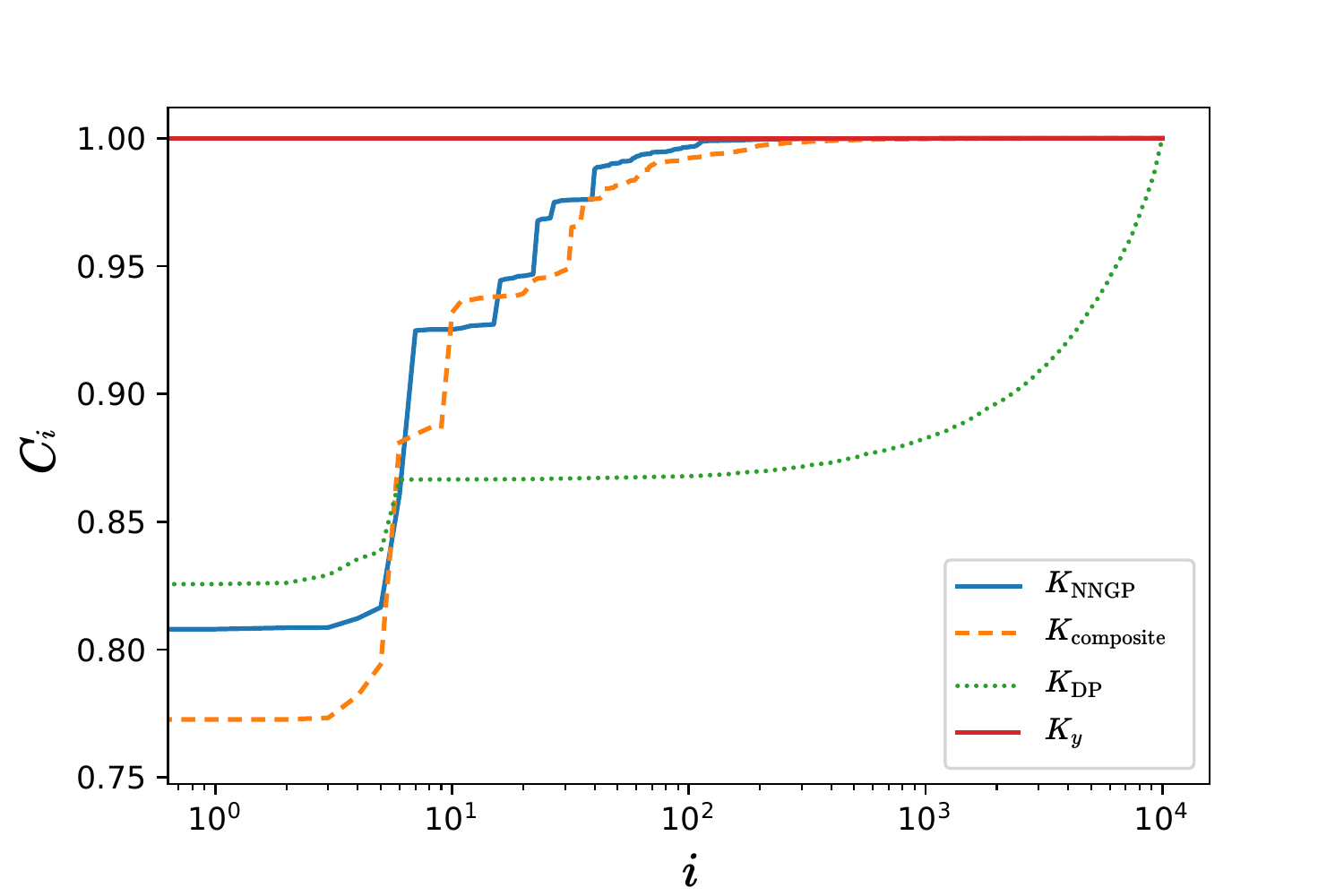} 
\caption{The cumulative power distributions for mode $i$ for models with different kernels: the solid curve -- NNGP model with $L = 5$; the dashed curve -- the composite kernel with $L = 5$; the dotted curve -- the dot-product kernel. All models are trained using 1000 randomly selected energy points. To obtain the CDs shown, the covariance matrix $K$ is computed using a distribution of 10000 randomly selected energy points excluding the training points. $K_y = \bm y^T \bm y$ is computed with the same distribution of 10000 points and is represented by the horizontal solid line.}
\label{fig:cumulative}
\end{figure}

As discussed in Section 2, the efficiency of a kernel regression model can be quantified by the CD $C(i)$. Figure \ref{fig:cumulative} compares the convergence of $C(i)$ for the NNGP model, the GP model with a composite kernel and the GP model with a dot-product kernel $k(\bm x, \bm x') = \bm x^T \bm x'$. 
For this calculation, the composite kernel is selected by maximizing BIC using the algorithm of Fig. 1 with $L=5$.  All kernel parameters are selected by maximizing the logarithm of the marginal likelihood with 1000 randomly selected energy points.  
When the kernel parameters are identified,  the kernel matrix $K$ is built using a set of randomly selected 10000 energy points excluding the training points.
The results show that NNGP models require fewer eigenvalues to converge $C(i)$ to one, indicating better alignment of the NNGP kernel function with the target function. Interestingly, we observe a similar rate of convergence for the NNGP model and the GP model with the composite kernel.

\begin{figure}[http]
\centering
\includegraphics[scale=1]{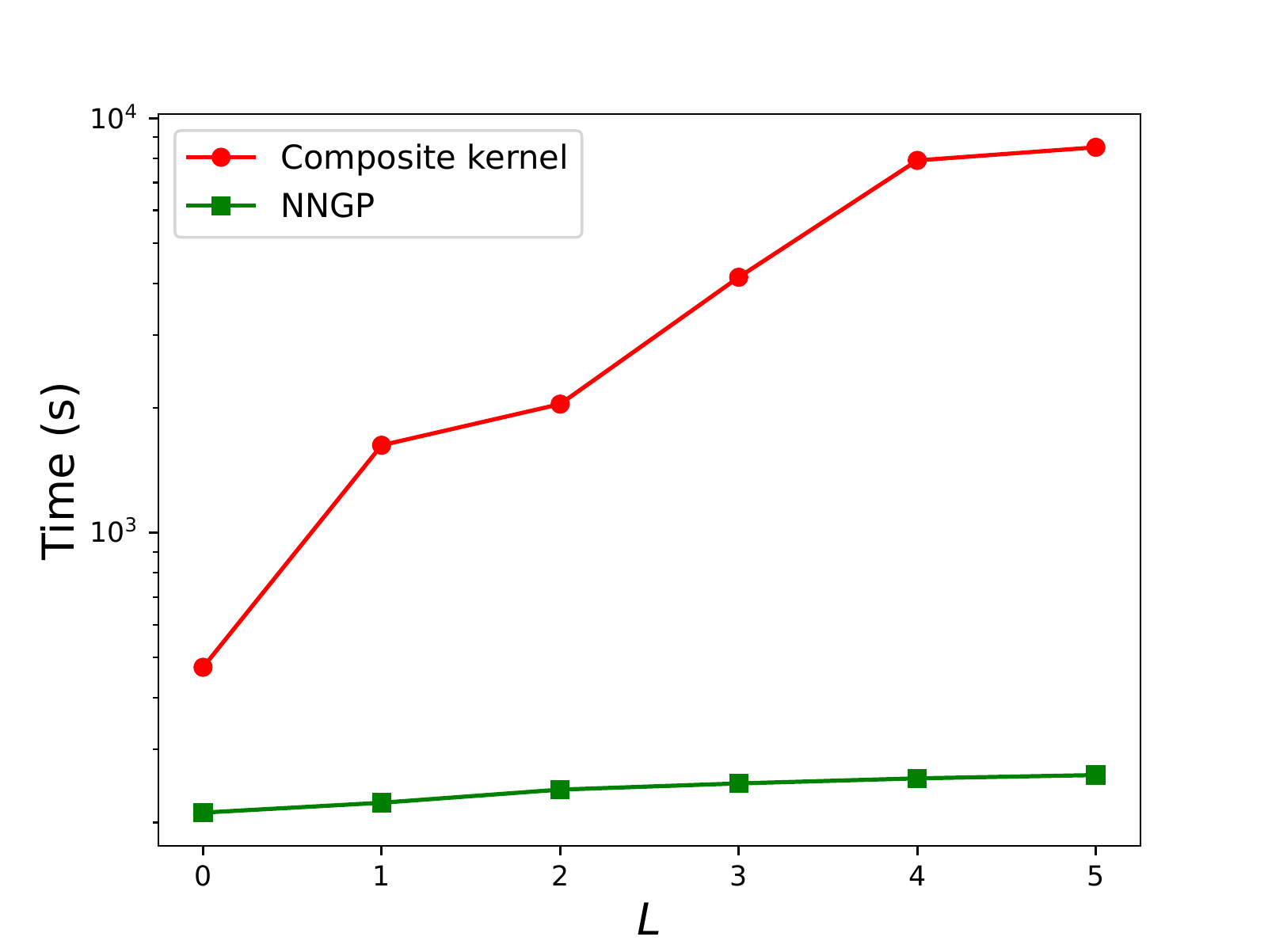} 
\caption{Comparison of the computation time required to train composite kernel (circles) and NNGP (squares) models. For the composite kernel, $L$ is the number of searched layers in the algorithm shown in Figure 1.  For the NNGP kernel, the computation time is recorded for training an NNGP kernel with $L$ layers. All models are trained using a fixed set of 1000 randomly selected energy points.}
\label{fig:time}
\end{figure}

The above results illustrate that NNGP models provide more accurate and data-efficient models of PES than GP models with composite kernels. At this same time, the computational complexity of training NNGP models is much lower than for GP models with composite kernels. To illustrate this, we plot in Figure \ref{fig:time} the computation time for training the two types of models as a function of the number of layers $L$ in the kernel construction algorithm. 
These calculations are performed on a computer with an AMD 5950x processor with 32GB of memory and an RTX3090 graphics card with 24GB of memory. The computation time for training the NNGP kernel does not significantly vary with $L$, due to the availability of analytical solutions and the use of GPU for computation. However, for the composite kernel, the computation time increases dramatically with $L$. The composite kernel with $L=5$ layers requires more than 39.2 times the amount of time needed to train the NNGP kernel.

\subsection{Free energy surface for Heisenberg model}

In order to further demonstrate the generalization accuracy of NNGP models for extrapolation tasks, we consider a different problem: the evolution of the free energy with temperature ($T$) and magnetization ($m$) for the one-dimensional (1D) Heisenberg model

\begin{eqnarray}
\hat H=-\frac{J}{2} \sum_{\langle i, j \rangle} \hat{S}_i \cdot \hat{S}_j,
\label{Heisenberg}
\end{eqnarray}

\noindent
where $J$ is the strength of the coupling between spins in different sites of an infinite one-dimensional lattice, $\hat S_i$ is the spin-$1/2$ operator for lattice site $i$ and the angular brackets indicate summation over indices of adjacent sites only. 
The dimensionless magnetization $m$ is defined as $2\left\langle {S} \right\rangle$,  where $\left\langle {S} \right\rangle$ is the average over all spin orientations in the lattice. 
The free energy density for this model can be computed using the mean-field approximation, yielding \cite{qt,qt-2,qt-3,qt-gp}

\begin{equation}
f(T, m) \approx \frac{1}{2}\left(1-\frac{T_c}{T}\right) m^2+\frac{1}{12}\left(\frac{T_c}{T}\right)^3 m^4.
\label{free_energy}
\end{equation}

\noindent
The result is a combination of a quadratic term in $m^2$ and a quartic term in $m^4$, which reflects the competition between the ferromagnetic interaction, inducing the alignment of spins, and the thermal energy, which tends to induce disorder in spin alignment. 
The system undergoes a phase transition from the paramagnetic phase to the ferromagnetic phase at temperature $T_c = J/4$.  In the paramagnetic phase, the spins are randomly oriented and the magnetization is zero, while in the ferromagnetic phase, the spins are aligned and the magnetization is non-zero. When $T > T_c$, the thermal energy is sufficient to overcome the ferromagnetic interaction and the system is in the paramagnetic phase. When $T < T_c$, the ferromagnetic interaction is stronger and the system is in the ferromagnetic phase. This is reflected by the evolution of the free energy density (\ref{free_energy}) from a single-well dependence on $m$ at $T > T_c$ to a double-well dependence on $m$ at $T < T_c$. Our aim is to explore if a NNGP model can be trained by the free energy density evolution in the paramagnetic phase to predict the critical temperature $T_c$ and the free energy density in the ferromagnetic phase by extrapolation along the temperature axis $T$. 

\begin{figure}[http]
\centering
\includegraphics[scale=0.51]{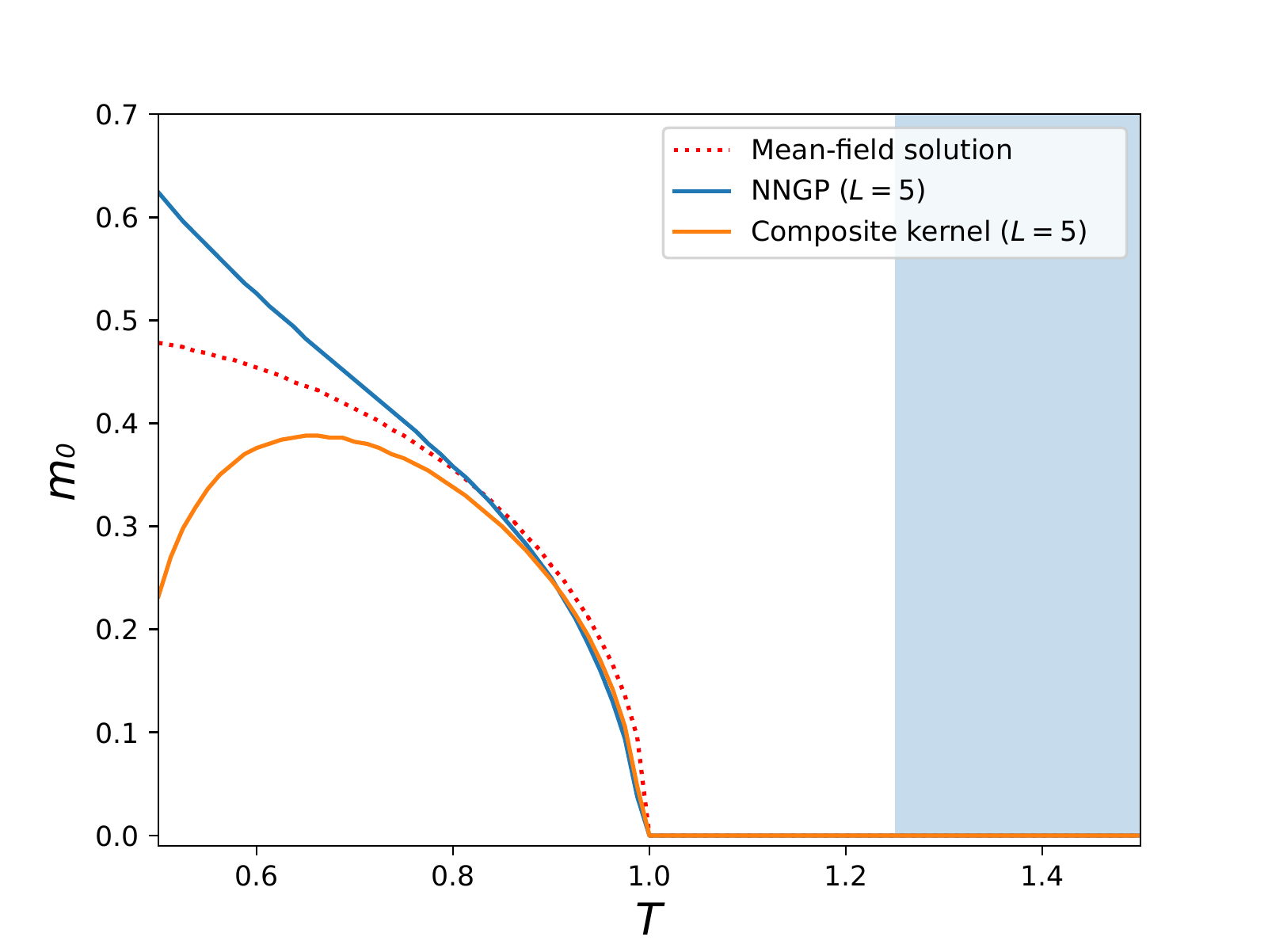} 
\includegraphics[scale=0.15]{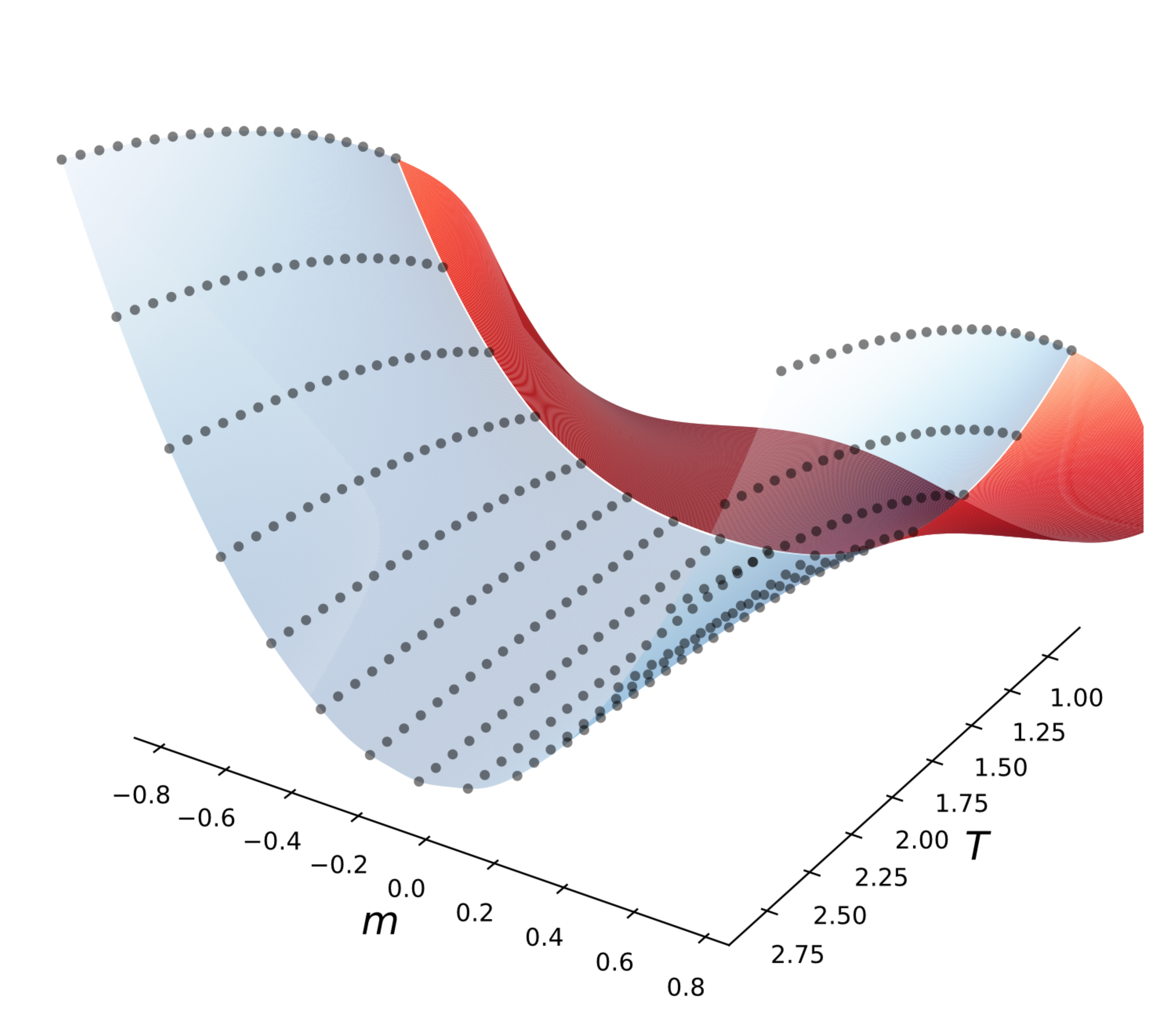} 
\caption{Left: The prediction of the order parameter $m_0$ that minimizes the free energy density $f(T, m)$ of the mean-field solution of the Heisenberg model by the NNGP model (upper solid curve) and GP model with composite kernel (lower solid curve).  
The mean-field result is shown by the dotted curve. 
The critical temperature of the phase transition is $T_c = 1$, and the training dataset consists of 300 points shown by symbols in the right panel. The training data are in the range of $T \in [1.25, 3]$ and $m \in [0,1]$.
Right: The surface $f(m, T)$ produced by the NNGP model with $L = 5$ by interpolation (blue area including the distribution of training points shown by symbols) and extrapolation (red area). 
}
\label{fig:heisenberg}
\end{figure}

We train two-dimensional models with $T$ and $m$ as input variables using 300 values of $f(m, T)$ given by Eq. (\ref{free_energy}) at temperatures in the shaded region of Figure \ref{fig:heisenberg} (left) and shown by the symbols in  Figure \ref{fig:heisenberg} (right). The training data are purposely far-removed from the critical temperature $T_c$. 
Figure \ref{fig:heisenberg} (left) compares the order parameter 
\begin{eqnarray}
m_0(T) =  \text{arg}\,\min\limits_{m}\ f(m, T),
\end{eqnarray}

\noindent
predicted by the NNGP model with $L=5$, the GP model with the composite kernel with $L = 5$ and the mean-field results (\ref{free_energy}). 
Both GP models provide an accurate prediction of $m_0$ deep into the ferromagnetic phase without any information from the ferromagnetic phase. 
Figure \ref{fig:heisenberg} (right) depicts the surface of $f(m,T)$ produced by interpolation (blue area including the distribution of training points shown by symbols) and extrapolation (the remaining red area) with the NNGP model. This figure illustrates that the NNGP models produce a physical, smooth surface by both interpolation and extrapolation in the input variable space.

\section{Conclusion}

The generalization accuracy of GP models can be enhanced by optimizing the complexity of the GP kernel function. There are two general methods for building GP kernels with variable complexity. GP models with composite kernels can be constructed by combining simple kernel functions into products and linear combinations using an iterative algorithm guided by Bayesian information criterion. Previous work showed that this algorithm of increasing kernel complexity can be used to enhance the accuracy of PES models for polyatomic molecules trained by a fixed distribution of a small number of energy points. However, building composite kernels by this iterative algorithm can be computationally expensive, for two reasons. First, the iterative kernel construction requires one to build multiple GP models with varying kernel function complexity.  Second, 
the number of kernel parameters increases quickly with the depth of the kernel composition search (as illustrated in Figure 1), making model training extremely time-consuming. 

In the present work, we consider an alternative algorithm of increasing kernel complexity for building GP models of PES with enhanced generalization accuracy. This algorithm exploits the connection between Bayesian NNs and Gaussian processes to yield NNGP models. 
We have shown that NNGP models yield PES with similar or better accuracy than GP models with composite kernels, at a fraction of the computation cost. 
To illustrate the generalization accuracy of the NNGP models, we consider two examples: a 6D PES for the molecular ion H$_3$O$^{+}$ and a 2D free energy density surface of the Heisenberg spin model. 
We demonstrate that 6D models trained by random samples from low energy distributions produce accurate PES both at low energies and at high energies, illustrating the ability of NNGP models to extrapolate in the energy domain. 
The 2D models are used to illustrate the extrapolation by NNGP models in the input variable space. 
We observe that the NNGP models are less sensitive to the distributions of training points and the model hyperparameters, yielding fast convergence with the number of NN layers. 
This makes NNGP models particularly suitable for use as surrogate models in Bayesian optimization for finding minima of PES that are expensive to compute. 
The improved computational efficiency of NNGP models, due in part to GPU acceleration, is expected to allow for modelling of complex high-dimensional systems.

\section*{acknowledgment}
This work was supported by NSERC of Canada

\clearpage
\newpage

\bibliographystyle{achemso}
\bibliography{refs}

\newpage


\end{document}